# Explanation of quantum dot blinking

# without long-lived trap hypothesis

**Pavel A. Frantsuzov**

Chemistry Department

University of California, Irvine

Irvine, CA 92697

**R. A. Marcus**

Noyes Laboratory of Chemical Physics

California Institute of Technology

Pasadena, CA 91125

(July 28, 2005)

A simple model explaining the experimental data on QDs luminescence blinking is suggested. The model does not assume the presence of the long-lived electron traps. The bleaching of the QD luminescence is a result of the Auger assisted radiationless relaxation of the excitation through the deep surface states. Possible ways of the experimental verification of the model are discussed.

# I. INTRODUCTION

Modern technology of colloidal nanocrystal preparation has revolutionized the field. This technology allows the fabrication of spherical semiconductor quantum dots (QD) with narrow size distribution.[1,2] Colloidal quantum dots having wide excitation spectra, narrow emission spectra, high quantum yields for luminescence, and great photostability are ideal laser material,[3,4] single-photon source,[5] and luminescent labels for chemical and potential biological applications.[6]

Investigations of the QDs properties by various experimental techniques have yielded a number of striking results. One of them is the statistics of the single QD luminescence switching events (blinking).[7] It is known from many single-molecule experiments that under continuous excitation the luminescence emission switches "on" and "off" by sudden stochastic jumps. Such a behavior was observed in dye molecules,[8] light harvesting complexes[9] and polymers.[10] In these non-QD systems the distribution of the "on" and "off" times is usually exponential or near-exponential.

In contrast, the distribution of the blinking events for the colloidal QDs has the form of a power law.[11-18] Single CdSe QD experiments conducted by Shimizu et al.[11] demonstrated the power law distribution for the both "on" and "off" intervals over the four decades of time, namely from 0.1s to 1000 s. The exponent was about –1.5 for the both distributions. This power law is unchanged for the different temperatures (from 10 K to room temperature), radii (15 Å and 25 Å), laser intensities (from 100 W/cm$^2$ to 700 W/cm$^2$), and atmospheric conditions (from air to vacuum).[11] The distribution is also the same for uncapped CdSe QD and for CdSe QD capped by several ZnS monolayers.[11] Analogous results were found by other groups[12-18] for various sizes (from 15 Å to 27 Å), materials (CdSe, TeSe, InP), and laser intensities (from 100 W/cm$^2$ to 20 KW/cm$^2$). The exponents for the "off" time distribution is in the range from -1.5 to -1.6 are reported for all the experiments. The longest time range was covered in experiments of Kuno et al. on single 27 Å radius CdSe QDs capped by ZnS at room temperature.[12,13] In the latter case the power law distribution with the exponent ~-1.5 was found to apply over more than five decades of time, namely from 200 μs to 100 s. The power-law "on" times distribution with the exponent varying from about -1.5 to –2.0 was observed by different experimental groups.[12-18] In the present article we address mostly to the results of Shimizu et al.[11] where the exponent around –1.5 was observed for the "on" and "off" times distributions.

A standard explanation of blinking phenomenon is given by a three-level model.[19] This model has a ground state, a light-emitting excited state, and a "dark" trapping state from which the system does not luminesce. A laser is used to excite the system from the ground state to the light-emitting state. A luminescence occurs with the emission of a photon from the light-emitting excited state. Occasionally the light-emitting state undergoes a radiationless transition to a "dark" trapping state. Thereby, trapping and detrapping events switch the photoluminescence "off" and "on", respectively.

A now commonly accepted idea on the nature of the trapping state in the colloidal QDs was proposed by Efros and Rosen (a long-lived trap hypothesis).[20] They suggested that the luminescence is quenched if one of the carriers (the electron or the hole) is trapped in the surrounding matrix. A photoexcited electron-hole pair in such a charged dot can recombine by a radiationless Auger process. The recombination time of the Auger process, about 1 ps,[21,22] is three or four orders of magnitude faster than a radiation assisted process, and so the QD then has an extremely small probability to emit a photon after excitation. The charged QD is "dark". This mechanism of the luminescence quenching is supported by several experimental observations: qualitative quenching of the luminescence is found after the addition of an electron[23] or hole quencher, [23,24] thus making this charged QD dark.

The model of Efros and Rosen[20] connects the luminescence switching events with the electron transitions from the QD to the long-lived trap state and back. A simple three-level picture with constant ionization and neutralization reaction rates did not, however, explain the power law distribution of the blinking times. This model gives an exponential distribution of "off" and "on" times. [20] More advanced explanations including multiple traps, like a model suggested in Ref. 25, could in principle explain the power law distribution of "off" times, but not "on" times, as was discussed in details by Kuno et al.[5,26]

Three major models of the power law behavior utilizing the long-live trap hypothesis have been suggested in the literature. Kuno et al. [26] gave an explanation of the power law distribution using a wide range of exponentially distributed ionization-recombination rates, switched randomly after each electron transition between the internal QD states and the electron traps in the environment. The model assumes very large variations of the electron transition rate (5-6 orders of magnitude) caused by the changes in the environment. Such large fluctuations of the electron transfer rate, however, have not been observed in other systems.

Shimizu et al.[11] proposed a model of a resonant transitions of the electron between the excited state of the QD and a long-lived (with intrinsic lifetime more than 1000 s) trap state. It is known from the experiment[27-29] that the energy of the excited state displays a long-correlated stochastic motion with a characteristic time of hundreds of seconds. The amplitude of this spectral diffusion varies from 3 meV at 10K to 60 meV at room temperature.[27] At the cryogenic temperatures the diffusion is light- induced, and the shift of the luminescence line depends directly on the number of the absorbed photons. There is little published information[29] on the rate of the spectral diffusion at room temperature. According to the model of Shimizu et al.[11] the electron transfer event switching the luminescence intensity happens only when the excited state and the trap state are in resonance. The model provided a promising concept of a slow diffusive coordinate. The excitation energy of the lowest excited electronic state of the QD plays a role of this coordinate. The idea of the slow coordinate switching the intensity of the luminescence naturally explains the large difference between the excitation-relaxation times of the electronic system of the QD and the blinking times. Further development of this idea was made in the article of Jung et al.[30] The weak point in this model is an assumption of an existence of a single electron trap state placed in the narrow region of energies and located in the surrounding near each given QD. In order to resolve this problem Tang and Marcus[31] assumed that the trap is the crystal-induced surface state. In the metals such states were determined by angle resolved photoemission and inverse photoemission of valence-edge and conduction-edge states.

The third model proposed by Margolin et al. [32] gives an explanation in which the blinking by 3-dimensional hopping diffusion of the photoejected electron in the surrounding media. The positively charged QD stays "off" until the electron returns back. The long "on" times are explained by the existence of a long-lived hole trap in the vicinity of the QD. While the hole is trapped and the electron is diffusing the QD stays "on". A weakness of this model is in non-zero escape probability of the ejected electron.

We explore here an alternative mechanism for the luminescence intermittency in which does not assume any long-lived trap state. The QD always returns back to the ground (neutral) state after photoexcitation ( directly or via a surface state ). The "on"-"off" switching of the QD luminescence intensity is caused by large variations of the nonradiative relaxation rate of the excited electronic state to the ground state via surface hole trap states. The hole trapping is assuming to be induced by an Auger-assisted mechanism.

The experimentally measured electric, optical, mechanical and kinetic properties of the quantum dots given in this article concern CdSe colloidal QDs only.

II. MECHANISM OF THE QD INTERMITTENCY

The size of the colloidal QDs used is smaller than a size of the Bohr exciton in the bulk (~112 Å for CdSe[33]). So, they are in the so-called strong confinement regime, where the electrons and holes could be considered as independent particles, taking into account the Coulomb interaction as a perturbation.[33] The structure of the delocalized electronic levels in the QD could be described using simple particle-in-the-box model (see Fig.1.). The levels in the conduction band correspond to the single electron states in the spherical well. For the electron states with the envelope angular momentum 0, 1, 2, …, we use notations $1S_e$, $1P_e$, $1D_e$, etc., respectively with a subscript e. The valence band levels correspond to the single hole states ($1S_{3/2}$, $1P_{3/2}$, $2S_{3/2}$, etc.), the subscript denoted the total angular momentum which is the sum of the envelope angular momentum and the band-edge Bloch function angular momentum.[33] The uncharged QD in the ground state contains no electrons or holes. The photoexcitation of the QD creates an electron in the conduction band and a hole in the valence band. The lowest excitation energy of the QD is calculated as

$$E_{ex} = E_{1Se} - E_{1S3/2} + U(1S_e, 1S_{3/2}) \qquad (2.1)$$

where $E_{1Se}$ and $E_{1S3/2}$ are the energies of $1S_e$ and $1S_{3/2}$ single particle levels and $U(1S_e, 1S_{3/2})$ is the energy of the averaged Coulomb interaction between the electron in $1S_e$ and the hole in $1S_{3/2}$ state[33]

$$U(1S_e, 1S_{3/2}) \approx -1.8 \frac{e^2}{\kappa R} \qquad (2.2)$$

where $e$ is the electron charge, $R$ is a radius of the dot and $\kappa$ is a dielectric permittivity of the semiconductor material. For the CdSe QDs with the radius 15-35 Å the excitation energy $E_{ex}$ given by Eqs. (2.1) - (2.2) varies from 2.1 eV to 2.6 eV.

The emission spectrum of the CdSe QD photoluminescence has a narrow peak at the energy $E_{ex}$. There is also a broad and less intense luminescence maximum at 300-700 meV lower than the excitation energy,[34,35] presumably due to the electron or hole transition into the surface states in the QD. It was demonstrated that the surface state emission intensity is very sensitive to the removal and the exchange of the ligand layer.[36] Optically detected magnetic resonance studies have showed that this maximum corresponds to deep hole traps.[37] We can conclude that the surface states are placed lower than the Fermi level and 300-700 meV higher than the $1S_{3/2}$ state. A broad surface states emission peak with the energy 400 meV lower than $E_{ex}$ was found by an electrogenerated chemiluminescence measurements for ZnS capped CdSe QDs[38] as well as for uncapped ones.[39] A similar placement of the surface states was obtained by a scanning force conductance spectroscopy.[40] Numerical simulations of the QD electronic structure demonstrate that the trap states could form a band.[41] Formation of the surface states band has a reasonable explanation. The 15 Å radius CdSe QD consists of about 1000 atoms, more than 400 of them existing on the surface. So, there are at least 400 dangling bonds in the QD. It was demonstrated[36, 42] that only Cd atoms on the surface (but not all of them) are connected to the TOPO ligands. As a result, we have at least 200 surface states in the band-gap corresponding to Se dangling bonds. The corresponding electronic states weakly interact with each other and as a result form the band. We assume that the hole trap states in any given QD form a band with the width of about 200 meV (see Fig.1). The energy gap between $1S_{3/2}$ state and the lowest level in the band $E_1$ could vary from dot to dot, but it is always about 300-400 meV.

Absorption spectroscopy of the negatively charged QDs shows a peak in the infrared region corresponding to $1S_e \rightarrow 1P_e$ transition of the excess electron in the conduction band.[43] The center of the peak $\varepsilon_0$ corresponds to the energy difference between $1S_e$ and $1P_e$ states predicted theoretically[33] (~300 meV for 24 Å radius QD). The full width at half maximum (FWHM) of the peak varies from 140 meV to 220 meV when the QD radius changes from 26 Å to 15Å. The infra-red absorption spectroscopy of the QD ensemble after the visible excitation also shows a peak[44,45] centered at the same energy with a similar width ( ~170 meV for 23 Å radius QD). The absorption line rises in few picoseconds after the visible excitation and decays very slowly, at times of $\tau_n \sim 1$ µs.[45] It is the clear evidence that fast hole trapping occurs in a large fraction of the QDs in the ensemble. While the hole is trapped, the electron is located in $1S_e$ state in the conduction band and can be excited to $1P_e$ state by IR irradiation. This peak in the IR absorption spectrum could not be attributed to the electron trapping, because the hole excitation energy is much smaller due to large effective mass. The slow decay of the absorption corresponds to a non-radiative phonon– assisted recombination of the trapped hole and the electron through the large energy gap. That wide spectrum of $1P_e$-$1S_e$ energy difference

$$\varepsilon = E_{1Pe} - E_{1Se} \qquad (2.3)$$

of about 200 meV is difficult to explain by the size distribution of the QDs in the ensemble. The energy difference $\varepsilon$ scales as $R^{-2}$ with the radius of the dot $R$.[33] Reported size distribution (about 5% rms) of the QD ensemble could so generate inhomogeneous FWHM not larger than 70 meV. On the other hand, hole-burning experiments on this $1P_e$-$1S_e$ absorption demonstrate a quite narrow homogeneous line.[46] Thus, we suggest that the energy difference between $1S_e$ and $1P_e$ electron states ε is a subject of a light-induced diffusion, analogous to the diffusion of the luminescence line energy $E_{ex}$. Specifically, we consider the difference of the energies of the electronic states ε as a function of the nuclear coordinates of the system (QD + ligand layer). The small movements of the nuclear coordinates after the photoexcitation of the dot is assumed generate a slow light-induced diffusion of this variable. Subtracting the calculated inhomogeneous FWHM from the observed width we have an estimate ~100 meV for the diffusion amplitude of ε. We discuss possible mechanisms of the light –induced spectral diffusion later.

  As is seen from experiment,[35] the deep trap emission rises in few picoseconds after the light excitation demonstrating surprisingly high trapping rate for such a large energy gap. The overlap integral between the delocalized $1S_{3/2}$ state and the localized surface state has to be very small in comparison with the overlap of the delocalized states. So, the phonon-assisted hole trapping process is assumed to be very slow in this case. We suggest that the trapping is an Auger–type resonant process (see Fig.2): the excess hole energy provides to the electron excitation.

The kinetic scheme of the excitation-relaxation cycle of the QD is presented in Fig.2. The scheme is described by a system of kinetic equations

$$\frac{d}{dt}P_0 = -k_I P_0 + k_r P_e + k_n P_t$$

$$\frac{d}{dt}P_e = k_I P_0 - (k_r + k_t)P_e + k_d P_i$$

$$\frac{d}{dt}P_i = k_t P_e - (k_d + k_e)P_i \qquad (2.4)$$

$$\frac{d}{dt}P_t = k_e P_i - k_n P_t$$

where $P_0, P_e, P_i,$ and $P_t$ are the populations of the ground state, the excited state, the intermediate state (trapped hole + $1P_e$ electron) and the trap state (trapped hole + $1S_e$ electron), respectively. The populations satisfy the normalization condition

$$P_0 + P_e + P_i + P_t = 1 \qquad (2.5)$$

The excitation rate is expressed as[12]

$$k_I = \frac{I\sigma}{\omega} \qquad (2.6)$$

where $I$ and $\omega$ are the intensity and the frequency of the light and $\sigma$ is the excitation cross-section. The characteristic laser intensity $I = 1$ kW/cm$^2$ corresponds to an excitation rate $k_I \sim 10^7$ s$^{-1}$, using $\sigma \sim 4 \times 10^{-15}$ cm$^2$ for the absorption cross-section of the dot.[13] The radiative decay rate of the excitation $k_r$ is the inverse of the radiative lifetime $\tau_r$. The radiative lifetime of the QD excitation is about 150 ns at 10 K.[34] It decreases with increasing temperature and reaches the value ~20 ns at room temperature.[47,48] Such dependence is attributed to the splitting in the fine structure of the excited state (existence of a dark exciton).[33,47] We set $\tau_r$ equal to 20 ns in our model. The population probability of the excited state is always much less than unity since $k_I \ll k_r$. So the effect of a double excitation and of an induced emission can be neglected. The rate of the phonon-assisted 1P$_e$ → 1S$_e$ transition without the hole excitation (non-Auger process) $k_e$ was found[49] to be ~3×10$^{11}$ s$^{-1}$. The nonradiative relaxation of the trapped state (trapped hole + 1S$_e$ electron) to the ground state of the QD (the electron-hole recombination) $k_n = 1/\tau_n$ is in order of 10$^6$ s$^{-1}$, according to Ref. 45.

The rate of the Auger assisted trapping as a function of the reaction coordinate $\varepsilon$ defined by Eq.(2.3) is given by the Fermi Golden rule formula:

$$k_t(\varepsilon) = \frac{2\pi}{\hbar} V_{eh}^2 n\left(\varepsilon + E_{1S_{3/2}} + U(1P_e, TS) - U(1S_e, 1S_{3/2})\right) \quad (2.7)$$

where $V_{eh}$ is the matrix element of the electron-hole interaction, $n(E)$ is the density of the surface states as a function of energy and $U(1P_e, TS)$ is the averaged Coulomb interaction between the electron in 1P$_e$ state and the trapped hole. The calculated value[44] of $U(1P_e, TS)$ slightly varies from $-1.15 e^2/\kappa R$ to $-0.925 e^2/\kappa R$ depending on the position of the trapped hole on the surface. The fast 1P$_e$ → 1S$_e$ relaxation generates a finite width $\Gamma = \hbar k_e \sim 200$ µeV for the 1P$_e$ state energy. In theory, the energy $E_{1Pe}$ could be exactly determined in the absence of the electron-phonon interaction. We use this "bare" value of the 1P$_e$ state energy for the precise definition of the variable $\varepsilon$ in Eq. (2.3). According to our previous suggestion the density of surface states $n(E)$ consists of many peaks at the energies $E_1$, $E_2$, $E_3$, $E_4$ ... with spacing of about 1 meV. The small lifetime of the final state 1P$_e$ broadens each peak in Eq. (2.7) up to the width $\Gamma$. The standard theory predicts well-known Breit-Wigner shape of the decaying level. It was demonstrated however that the shape of the single level connected to the quantum chaotic system transforms into the Gaussian form in so-called strong coupling limit.[50] In this case the trapping rate could be cast as

$$k_t(\varepsilon) = \sum_i A_i \exp\left(-\frac{(\varepsilon - \varepsilon_i)^2}{\Gamma^2}\right) \qquad (2.8)$$

where $\varepsilon_i$ is expressed via the energy $E_i$ of the i-th surface state in the band.

$$\varepsilon_i = E_i - E_{1S_{3/2}} - U(1P_e, TS) + U(1S_e, 1S_{3/2})$$

The behavior of the trapping rate (2.8) is illustrated in Fig.3, when the rate increases many orders of magnitude in the vicinity of $\varepsilon_1$. We estimate the value of $\varepsilon_1$ as 300 meV. There are no experimental data on the detrapping rate $k_d$. Due to the similarity of the molecular mechanism we assume that the detrapping rate is of the same order of magnitude as the trapping rate $k_t$. The exact numerical value of this rate does not affect the luminescence properties, as seen later. The luminescence intensity $w$ the proposed model, given in Eq. (2.4), is proportional to the excited state population $P_e$

$$w = k_r P_e \tag{2.9}$$

We note that while the system is trapped in the surface state, another electron-hole pair can be excited by the photon absorption. This pair, however, quickly recombines by Auger mechanism and doesn't add to the luminescence intensity. Thus, we do not consider those excitations in Eq. (2.4). The Eqs. (2.4), (2.5), and (2.9) give the following expression for the steady state luminescence intensity under the continuous laser excitation

$$w = \frac{k_I}{A + \alpha_t B} \tag{2.10}$$

where

$$\alpha_t = \frac{k_t}{k_d + k_e}, \quad A = 1 + \frac{k_I}{k_r}, \quad B = \frac{k_e}{k_r} + \frac{k_I}{k_r} + \frac{k_I}{k_r}\frac{k_e}{k_n}$$

Using estimate for the reaction rates ($k_I \sim 10^7$ s$^{-1}$, $k_r \sim 5 \times 10^7$ s$^{-1}$, $k_e \sim 3 \times 10^{11}$ s$^{-1}$, $k_n \sim 10^6$ s$^{-1}$) we have

$$A \sim 1 \quad ; \quad B \sim 6 \times 10^4$$

It follows from Eq. (2.10) that the emission intensity is about $k_I$ when $k_t$ is much less than $k_t^* = k_e / B \sim 5 \times 10^6$ s$^{-1}$ (the luminescence is "on"). In opposite case $w$ in Eq. (2.10) is much less than $k_I$ (the luminescence is quenched). It is readily seen that the result does not depend on $k_d$ in both limiting cases. It could be omitted it in Eq. (2.10), yielding a simpler formula:

$$w = \frac{k_I}{A + k_t / k_t^*} \tag{2.11}$$

In the proposed model of the QD blinking the $1P_e$–$1S_e$ energy gap $\varepsilon$ plays a role of the slow variable. It performs the light-induced stochastic motion generating huge variations of the trapping rate. Such variations of the nonradiative relaxation rate of the excited QD were observed experimentally by Schlegel et al.[48] by time-correlated photon counting. The motion of the stochastic variable $\varepsilon(t)$ is much slower than the excitation-relaxation transitions of the QD, one can utilize a local steady-state approximation.[51] Accordingly, the luminescence intensity at any given time $t$ is calculated using a stationary solution (2.11), where $k_t$ is treated as a constant:

$$k_t = k_t(\varepsilon(t))$$

As seen from Fig. 3 there is some value $\varepsilon^*$ very close to $\varepsilon_1$ separating two regions with the small and the large trapping rates

$$k_t(\varepsilon) \ll k_t^* \text{ , for } \varepsilon < \varepsilon^* - \delta$$

$$k_t(\varepsilon) \gg k_t^* \text{ , for } \varepsilon > \varepsilon^* + \delta \quad (2.12)$$

with a narrow transition layer of size $\delta \sim 20\,\mu eV$ in between. As a result, the luminescence intensity (2.11) is a nearly step function of $\varepsilon$. It causes the telegraph-like behavior of the luminescence intensity as a function of time: the luminescence intensity makes sudden jumps between "on" and "off" regimes. Fig. 4 gives an illustration of the model. When $\varepsilon$ which was initially smaller than $\varepsilon^*$ crosses the border and become larger than $\varepsilon^*$, the luminescence switches from "on" to "off". It switches back to "on" when the trap energy become smaller than $\varepsilon^*$ again.

## III. DISTRIBUTION OF THE BLINKING TIMES

The probability distribution function of the coordinate $\varepsilon$ at the time $t$ satisfies the following diffusion equation

$$\partial_t \rho(\varepsilon,t) = D\partial_\varepsilon (\partial_\varepsilon + (\varepsilon - \varepsilon_0)/\Delta^2)\rho(\varepsilon,t) \qquad (3.1)$$

where $\Delta \sim 50$ meV is a root mean square deviation of the (Gaussian) steady-state distribution of $\varepsilon$, and $\varepsilon_0 \approx 300$ meV is the center of the distribution. The diffusion coefficient $D$ is assumed to be proportional to the excitation intensity $I$ and to depend only weakly on temperature. However it could depend on the environment and the QD preparation procedure. We introduce the difference $x = \varepsilon - \varepsilon_0$ as a new reaction coordinate. This $x$ is a function of the nuclear coordinates of the system mentioned earlier. The master equation for the distribution function $\rho(x,t)$ follows from Eq.(3.1)

$$\partial_t \rho(x,t) = D\partial_x (\partial_x + x/\Delta^2)\rho(x,t) \qquad (3.2)$$

The threshold coordinate value between "on" and "off" regions is $x^* = \varepsilon^* - \varepsilon_0$. When the edge energy $\varepsilon^* \sim 300$ meV is very close to $\varepsilon_0$; the following condition is fulfilled

$$|\varepsilon^* - \varepsilon_0| \ll \Delta$$

Under this condition one can set $x^* = 0$ without loss of generality.

The "off" time period started when the coordinate $x$ crosses the transition region and becomes equal to $\delta$. We set time $t$ equal to zero at this moment. We show later that the final result for the observed distribution of the "off" times does not depend on the precise value of $\delta$. The distribution of the "off" times $p_{off}(t)$ is a derivative of the survival probability of the "off" state

$$p_{off}(t) = -\frac{d}{dt} P_{off}(t) \tag{3.3}$$

The survival probability is equal to unity at zero time and goes to 0 at infinite time. It could be expressed as an integral of the survival probability distribution function $\rho_{off}(x,t)$

$$P_{off}(t) = \int_0^\infty \rho_{off}(x,t) dx \tag{3.4}$$

$\rho_{off}(x,t)$ satisfies the equation (3.2)

$$\partial_t \rho_{off}(x,t) = D \partial_x (\partial_x + x/\Delta^2) \rho_{off}(x,t) \tag{3.5}$$

with an absorbing boundary condition at the border (the first passage time problem)

$$\rho_{off}(x,t)\big|_{x=0} = 0 \tag{3.6}$$

The condition (3.6) is not exact because of the finite size of the transition region $\delta$ near $x = 0$. However, it is valid for times much larger than

$$\tau_0 = \frac{\delta^2}{D} \tag{3.7}$$

The initial value of the coordinate $x$ at zero time is taken equal to $\delta$.

$$\rho_{off}(x, t=0) = \delta(x - \delta) \tag{3.8}$$

The solution of Eqs. (3.5), (3.6) and (3.8) is well known[52]

$$\rho_{off}(x,t) = \rho_0(x,t) - \rho_0(-x,t) \tag{3.9}$$

where $\rho_0(x,t)$ is a solution of Eq. (3.5) with the initial condition (3.8), but without the boundary conditions (3.6)

$$\rho_0(x,t) = \frac{1}{\sqrt{2\pi\Delta^2(1-\exp(-2t/\tau))}} \exp\left(\frac{-(x-\delta\exp(-t/\tau))^2}{2\Delta^2(1-\exp(-2t/\tau))}\right) \tag{3.10}$$

where

$$\tau = \Delta^2/D \tag{3.11}$$

is the diffusion relaxation time. The survival probability of the "off" state can be expressed as

$$P_{off}(t) = \frac{1}{\sqrt{2\pi\Delta^2(1-\exp(-2t/\tau))}} \int_{-b}^{b} \exp\left(\frac{-x^2}{2\Delta^2(1-\exp(-2t/\tau))}\right) dx \qquad (3.12)$$

where $b = \delta \exp(-t/\tau) \ll \Delta$. At the times $t \gg \tau_0$ the integrand in Eq.(3.12) is equal to 1 for the region of integration. So, we have

$$P_{off}(t) = \frac{2\delta \exp(-t/\tau)}{\sqrt{2\pi\Delta^2(1-\exp(-2t/\tau))}}, \qquad \text{for} \quad \tau_0 \ll t$$

This expression has the following limiting behavior for $P_{off}(t)$

$$P_{off}(t) = \frac{\delta}{\Delta}\sqrt{\frac{\tau}{\pi t}}, \qquad \text{for} \quad \tau_0 \ll t \ll \tau \qquad (3.13a)$$

$$P_{off}(t) = \frac{\delta}{\Delta}\sqrt{\frac{2}{\pi}}\exp(-t/\tau), \qquad \text{for} \quad t > \tau \qquad (3.13b)$$

The "off" time distribution function Eq. (3.3) yields the power law behavior

$$p_{off}(t) = \frac{\delta}{2\Delta}\sqrt{\frac{\tau}{\pi}} t^{-3/2} \qquad (3.14a)$$

for $\tau_0 \ll t \ll \tau$ and the exponential decay when $t \gg \tau$

$$p_{off}(t) = \frac{\delta}{\tau\Delta}\sqrt{\frac{2}{\pi}}\exp(-t/\tau) \qquad (3.14b)$$

This behavior is natural for any model of 1D diffusion with an absorbing boundary. The distribution $\rho_{off}(x,t)$ at small times is much narrower than $\Delta$. So, an unbiased diffusion approximation could be used which gives the power law result (3.14a). The exponential dependence at large times in Eq. (3.14b) corresponds to the decay of the quasi-stationary distribution $\rho_{off}(x,t)$ as a whole. Specifically, there is a reactive flux proportional to the survival probability $P_{off}(t)$, generating the first order steady-state decay rate $1/\tau$.

The distribution (3.14a) depends on the excitation light intensity and the temperature through $\tau$. The experimentally observed "off" distribution, however, does show any excitation dependence. It is a consequence of a measuring procedure including the integration of the luminescence intensity during time bin. The experimentally observed time distribution is limited to times greater than some minimal observation time $\tau_{min}$, which exceed the experimental time bin; one cannot see a blink shorter than that time. In order to observe the power law on the short times $\tau_{min}$ has to be larger than the smallest time $\tau_0$ for which of the power law is valid, given by Eq. (3.7). A measured normalized "off" distribution $\bar{p}_{off}(t)$ is proportional to $p_{off}(t)$ for the times $t$ larger then $\tau_{min}$.

$$\bar{p}_{off}(t) = C\, p_{off}(t)$$

The coefficient C could be found from the normalization integral

$$\int_{\tau_{min}}^{\infty} \bar{p}_{off}(t)\, dt = C \int_{\tau_{min}}^{\infty} p_{off}(t)\, dt = 1 \qquad (3.15)$$

Under the condition

$$\tau_0 < \tau_{min} \ll \tau \qquad (3.16)$$

the main contribution to the normalization integral comes from the region of small times. The normalized "off" time distribution is then found from (3.15) to be

$$\bar{p}_{off}(t) = \frac{1}{2\tau_{min}} \left(\frac{\tau_{min}}{t}\right)^{3/2}, \quad \text{for} \quad \tau_{min} < t \ll \tau \quad (3.17a)$$

$$\bar{p}_{off}(t) = \left(\frac{\tau_{min}}{2\tau^3}\right)^{1/2} \exp(-t/\tau), \quad \text{for} \quad t > \tau \quad (3.17b)$$

The "off" time distribution has the power law form with exponential cut-off. The normalized power law distribution (3.17a) depends on $\tau_{min}$ only, and so is thereby independent of the light intensity and the temperature. It also does not depend on the transition layer size $\delta$. Combining Eqs. (3.7) and (3.11) we obtain the following expression for $\tau_0$ value

$$\tau_0 = \tau \left(\frac{\delta}{\Delta}\right)^2 \quad (3.18)$$

The cut-off time $\tau$ for the "off" times distribution found from the ensemble luminescence measurements[53] to be of order of $\tau \sim 1000$ s. The values of the transition layer size and the diffusion amplitude were estimated above as $\delta \sim 20\,\mu eV$ and $\Delta \sim 50$ meV. The substitution of those values to Eq. (3.18) gives the smallest time for the validity of the power law $\tau_0 \sim 100\,\mu s$. This value is of the same order of magnitude as minimal observation time $\tau_{min}$ in the experiments of Kuno et al.[12-14] and much smaller than $\tau_{min}$ in the experiments of Shimizu et al.[11] So, the present model is consistent at the small times.

The "on" time distribution also could be found from the Eq.(3.2). The "on" time period starts when the coordinate $x$ crosses the transition region. We again set $t = 0$ at this moment and introduce the "on" survival probability distribution $\rho_{on}(x,t)$ satisfying the equation (3.2) for negative $x$

$$\partial_t \rho_{on}(x,t) = D\partial_x(\partial_x + x/\Delta^2)\rho_{on}(x,t) \qquad (3.19)$$

with an absorbing boundary condition

$$\rho_{on}(x,t)\big|_{x=0} = 0 \qquad (3.20)$$

The initial condition is the delta function.

$$\rho_{on}(x,t=0) = \delta(x+\delta) \qquad (3.21)$$

Equation (3.19) with the conditions (3.20) and (3.21) is equivalent to the Eqs. (3.5), (3.6) and (3.8) after a substitution $x \to -x$. As a result we have the same distribution (3.17) for the measured "on" time

$$\bar{p}_{on}(t) = \frac{1}{2\tau_{min}}\left(\frac{\tau_{min}}{t}\right)^{3/2}, \qquad \text{for} \qquad \tau_{min} < t \ll \tau \qquad (3.22a)$$

$$\bar{p}_{on}(t) = \left(\frac{\tau_{min}}{2\tau^3}\right)^{1/2}\exp(-t/\tau), \qquad \text{for} \qquad t > \tau \qquad (3.22b)$$

The Markovian nature of the master equation (3.2) determines an additional important property of the emission intensity. The consecutive blink events are not correlated in our model, because the coordinate $x$ has no memory about its prior evolution.

The present model explains the following experimental data on the colloidal QD luminescence blinking:

1. Distribution of the "on" and "off" luminescence durations under the continuous light excitation have an inverse power law form with an exponent of $\approx 1.5$ over six decades in time.[11]

2. The "off" time distribution is unchanged for the different temperatures and laser intensities.[11-18]

3. Consecutive blink events are not correlated. There is no correlation between the "off" time duration and the subsequent "off" time. Similarly, there is no correlation between consecutive "on" times and also between consecutive "on" - "off" and "off"-"on" times in agreement with experiments.[13,14]

4. The "on" and "off" time distributions are stationary in time. The statistics of the blinking times collected during frame segments of the full measurement time show the same power law distribution.[13]

5. The exponential cut-off time is predicted to be inversely proportional to the excitation intensity. It could also depend on the temperature, the environment and the procedure of the QD preparation.

## IV. LIGHT-INDUCED DIFFUSION OF THE QD STATES

Early spectroscopic experiments[34] showed that the emission spectrum of the colloidal QD ensemble is quite wide (about 80 meV). It was explained by the inhomogeneous broadening due to the QD size distribution (about 5% rms). Unexpectedly, it was found that the homogenous luminescence spectral line measured in the single QD experiments is also much broader than expected width $\hbar/\tau_r$, where $\tau_r$ is the radiative lifetime. Posteriori investigations[27-29] showed that the energy of the emitted photon $E_{ex}$ displays a long-correlated stochastic motion with a characteristic time of hundreds of seconds. The amplitude of this spectral diffusion (FWHM) is about 3 meV at 10K. The spectral diffusion is driven by the light absorption at this temperature. The mean square displacement of the spectral line directly depends on the number of absorbed photons.[29] Following Blanton et al.[28] we attribute this behavior to photoinduced changes at the nuclear coordinates at the nanocrystal surface. The changes could be the result of the mechanical (acoustic phonon) oscillations of QD shape after light excitation observed experimentally by Cerullo et al. [54] The changes in the nuclear coordinates at the surface shift the energies of $E_{1Se}$ and $E_{1S3/2}$ states. As a result, each excitation of the QD induces a small jump of the $E_{ex}$ value. The direction and the length of the jump are random. The characteristic length of the jump $a_E$ is much smaller than the amplitude of the spectral diffusion $\Delta_E$. The distribution of coordinate shifts at each jump has a bias depending on the value of the coordinate. The characteristic time interval between successive excitations is given by formula

$$\tau_I = 1/k_I \tag{4.1}$$

where the excitation rate constant $k_I$ is given by Eq. (2.6). According to the random walk theory,[55] the evolution of the coordinate $E_{ex}$ for the coarse-grain energies scale much larger than the jump length $a_E$ and the coarse-grain times scale much larger than the excitation time interval $\tau_I$ and so could be considered as a continuous diffusion with bias. The distribution function $\rho_E(E_{ex},t)$ of the energies thus satisfies a diffusion equation

$$\partial_t \rho_E(E_{ex},t) = D_E \partial_E (\partial_E + \varphi'(E_{ex})) \rho_E(E_{ex},t) \tag{4.2}$$

where the "diffusion coefficient" $D_E = a_E^2 / \tau_I$ is proportional to the excitation light intensity $I$ and where a unitless "potential" $\varphi(E_{ex})$ corresponds to the steady-state distribution of the emitting energies $\rho_{st}(E_{ex})$

$$\rho_{st}(E_{ex}) = \exp(-\varphi(E_{ex})) \tag{4.3}$$

If the steady-state distribution has a Gaussian form the function $\varphi(E_{ex})$ is parabolic.

$$\varphi(E_{ex}) = \frac{(E_{ex} - E_0)^2}{2\Delta_E^2} \tag{4.4}$$

where $E_0$ is a center of the steady-state distribution $\rho_{st}(E_{ex})$. It can be demonstrated that the diffusion model is consistent with the experimental data. The solution of Eqs. (4.2) – (4.4) with delta-function initial condition

$$\rho_E(E_{ex},0) = \delta(E_{ex} - E_{in})$$

is given by the well-known expression[55]

$$\rho_E(E_{ex},t) = \frac{1}{\sqrt{2\pi\Delta^2(1-\exp(-2t/\tau_E))}} \exp\left(\frac{-(E_{ex}-\overline{E}_{ex}(t))^2}{2\Delta^2(1-\exp(-2t/\tau_E))}\right) \tag{4.5}$$

where $\tau_E = \Delta_E^2/D_E$ is the relaxation time and

$$\overline{E}_{ex}(t) = (E_{in} - E_o)\exp(-t/\tau_E) + E_0$$

Eq. (4.5) also yields the time-dependence of the mean square displacement of $E_{ex}$

$$\langle(E_{ex} - \overline{E}_{ex}(t))^2\rangle = \Delta_E^2(1-\exp(-2t/\tau_E)) \tag{4.6}$$

Thus, the width of the luminescence line (FWHM) integrated over the time $t$ is given by the following expression

$$\text{FWHM} = \Delta\sqrt{8\ln 2\,(1-\exp(-2t/\tau_E))} \tag{4.7}$$

The observed time dependences of the luminescence line width the CdSe QD[29] at the temperatures 10 - 40 K are seen in Fig. 5 to be fitted by the function (4.7). As also seen there, the parameters for the spectral diffusion depend weakly on temperature in the given range. For the excitation intensity $I = 85$ W/cm$^2$ the averaged time between excitations is $\tau_I \sim 1$ μs.[13] The results of fitting allow the determination of an averaged length $a_E$ of

the single jump

$$a_E = \Delta_E \sqrt{\tau_I / \tau_E}.$$

Its value varies from 0.1 µeV to 0.3 µeV for the data presented in Fig. 5, when the temperature varies from 10 K to 40 K.

We suggest above that the energy difference between $1P_e$ and $1S_e$ states in $\varepsilon$, which is the function of the nuclear coordinates of the system, is also a subject of a light-induced diffusion with **a** mechanism analogous to the mechanism of the diffusion of the emission energy $E_{ex}$. The single jump size

$$a = \Delta \sqrt{\tau_I / \tau} \sim 1.5 \text{ µeV}$$

is much larger than $a_E$ (0.1 to 0.3 µeV). The probability density of the delocalized electron near the surface is larger for $1P_e$ state than for $1S_e$ state. Thus the sensitivity of the $1P_e$ energy to the surface modifications could be much larger than the $1S_e$ energy. It gives **a** possible explanation of the difference between $a$ and $a_E$. The sensitivity of the $1P_e$ state to the surface conditions could also make $a$ dependent on the environment.

## IV. DISCUSSION

We have suggested a simple model to explain the experimental data on QDs luminescence blinking**,** without assuming the existence of the long-lived electron or hole trap. The QD always returns back to the ground (neutral) state after photoexcitation. The main assumption of the model is that the bleaching of the QD luminescence is a result of the fast radiationless relaxation of the excitation by transition of the hole to the deep surface traps. The hole is trapped to the surface state by the Auger mechanism: the excess hole energy, due to an energetically favorable transition, enables the simultaneous electron excitation from $1S_e$ state to $1P_e$ state. So, the radiationless relaxation rate sharply depends on the energy gap between $1P_e$ and $1S_e$ states, which is the function of the nuclear coordinates of the system. We assume that $1P_e$ -$1S_e$ gap is the subject of the long-time diffusion induced by the light excitation. The present model explains the important properties of the QD blinking, but some features were left unexplained.

One of them is a large difference between the cut-off times for the "on" times and "off" times found in single QD experiments[11] and ensemble luminescence measurements.[47] The cut-off time for "off" distribution is about 40 times larger than for the "on" distribution at room temperature.[47] This fact could be explained if the mean jump length depends on the $1P_e - 1S_e$ energy gap $\varepsilon$. It is equivalent to the introduction of the $\varepsilon$ dependence of the diffusion coefficient in Eq.(3.1). Such dependence could be result of the large difference in the length of the jumps made with or without photon irradiation at room temperature. While the diffusion coefficient is reasonably larger in the "on" region in comparison with the "off" one, the cut-off time of the "on" time distribution in Eq. (3.22) would need to be much smaller than the corresponding "off" cut-off time in (3.17).

Another possible explanation of the cut-off times difference is in the influence of the competing trapping process, namely Auger-type excitation of the $1D_e$ electron state. It is natural to assume that the $1D_e$- $1S_e$ energy gap is also the subject of the light induced diffusion. Its motion should be correlated with the motion of the $1P_e - 1S_e$ gap. So, when the $1P_e - 1S_e$ energy gap $\varepsilon$ becomes small enough, the $1D_e$- $1S_e$ gap could reach the resonance with the upper edge of the trap energies band and quench the QD emission. It can be shown that this cut-off mechanism could apply for "on" times, but because of the positioning of the states could not apply to the "off" times.

While the exponent of the "off" times distribution found by the most of the experimental groups is always about –1.5, the "on" times distribution varies for the different experiments. [11-18] It seems that the properties of the "on" distribution are more sensitive to the environment conditions and to the QD preparation procedure. The power-law "on" time distribution with the exponent different from –1.5 could be explained in the present model by the strong $\varepsilon$ dependence of the diffusion coefficient in the "on" region. The development of the model in this direction is the subject of future investigation.

It is difficult to test directly one of the major assumptions of the present model, the diffusion of the $1P_e - 1S_e$ energy gap. An optical single QD spectroscopy seems unsuitable for this purpose, because the rate of the radiationless transition from 1P state to 1S state is many orders of magnitude larger than the rate of the radiative one. It is easier to check experimentally the light-induced nature of a driving force of the QD blinking. The present model predicts that the cut-off times of the "on" and "off" distributions are inversely proportional to the excitation intensity. There is some evidence of this behavior in literature. Shimizu et all.[11] demonstrated that the cut-off time of the "on" time distribution becomes smaller when the excitation intensity increases. The averaged "on" time obtained by Banin et al.[56] decreases with increasing excitation intensity. Systematic studies of the intensity dependence of the cut-off times are suggested in order to demonstrate the relevance of the light-induced mechanism of the luminescence intermittency.

We suggest another critical experiment for testing this point. One could switch off the laser irradiation during the single QD spectroscopic measurements and then switch it on after some macroscopic time. According to the present model the slow variable $\varepsilon$ doesn't move or moves little during this outage. So, if the luminescence was "on" before the outage it stays "on" after it (the same is true for "off" state). The correlation between the luminescence intensity before and after the laser outage can quite possibly be checked experimentally. Chung and Bawendi[53] have demonstrated that such a correlation occurs for the QD ensemble luminescence.

An important component of the present mechanism is the suggestion that radiationless relaxation goes via the hole trap state. That suggestion could be directly tested by the single QD near-infrared spectroscopy. According to the present model each single QD should demonstrate deep trap emission. The luminescence intensity in this spectral region should be switched by sudden jumps, correlated with the blinks in the visible spectrum. The population of the trap states is larger during "off" periods, so the deep trap emission should be more intense that time.

The time-resolved measurements of the relaxation rate of the excited state in the single QD could be done by the time-correlated photon counting procedure[48] together with the maximal likelihood estimation method.[57,58] The long-lived trap hypothesis predicts very fast (sub-picosecond) relaxation of the electronic excitation of the charged ("dark") quantum dot and slow (20 ns) radiative relaxation in opposite case. The suggested model of the radiationless relaxation by hole trapping allows intermediate relaxation rates. The observation of the relaxation rates larger than $10^8$ s$^{-1}$ and less then $10^{12}$ s$^{-1}$ could provide an additional test of the present theory.

**Acknowledgments**

We are pleased to acknowledge the support of the National Science Foundation and the Office of Naval Research. The authors would also like to acknowledge the support of P.A.F. by the James W. Glanville Fellowship in Chemistry at the California Institute of Technology.

We are very grateful to Louis E. Brus, Eli Barkai, Igor Goychuk, and Irina Gopich for fruitful discussions and to David J. Nesbitt, Moungi Bawendi, Philippe Guyot-Sionnest, Uri Banin, X. Sunney Xie, Victor Klimov, Masaru K. Kuno, Kentaro Shimizu, and Inhee Chung for the patience and kind explanations of the different aspects of the single-molecular spectroscopy and the QD blinking phenomenon in discussions with P.A.F.

34

56. U. Banin, M. Bruchez, A. P. Alivisatos, T. Ha, S. Weiss, and D. S. Chemla, J. Chem. Phys. **110**, 1195 (1999).

57. M. Köllner, J. Wolfrum, Chem. Phys. Lett. **200**, 199 (1992).

58. H. Yang, G. Luo, P. Karnchanaphanurach, T.-M. Louie. I. Rech, S. Cova, L. Xun, X.S. Xie, Science **302,** 262 (2003).

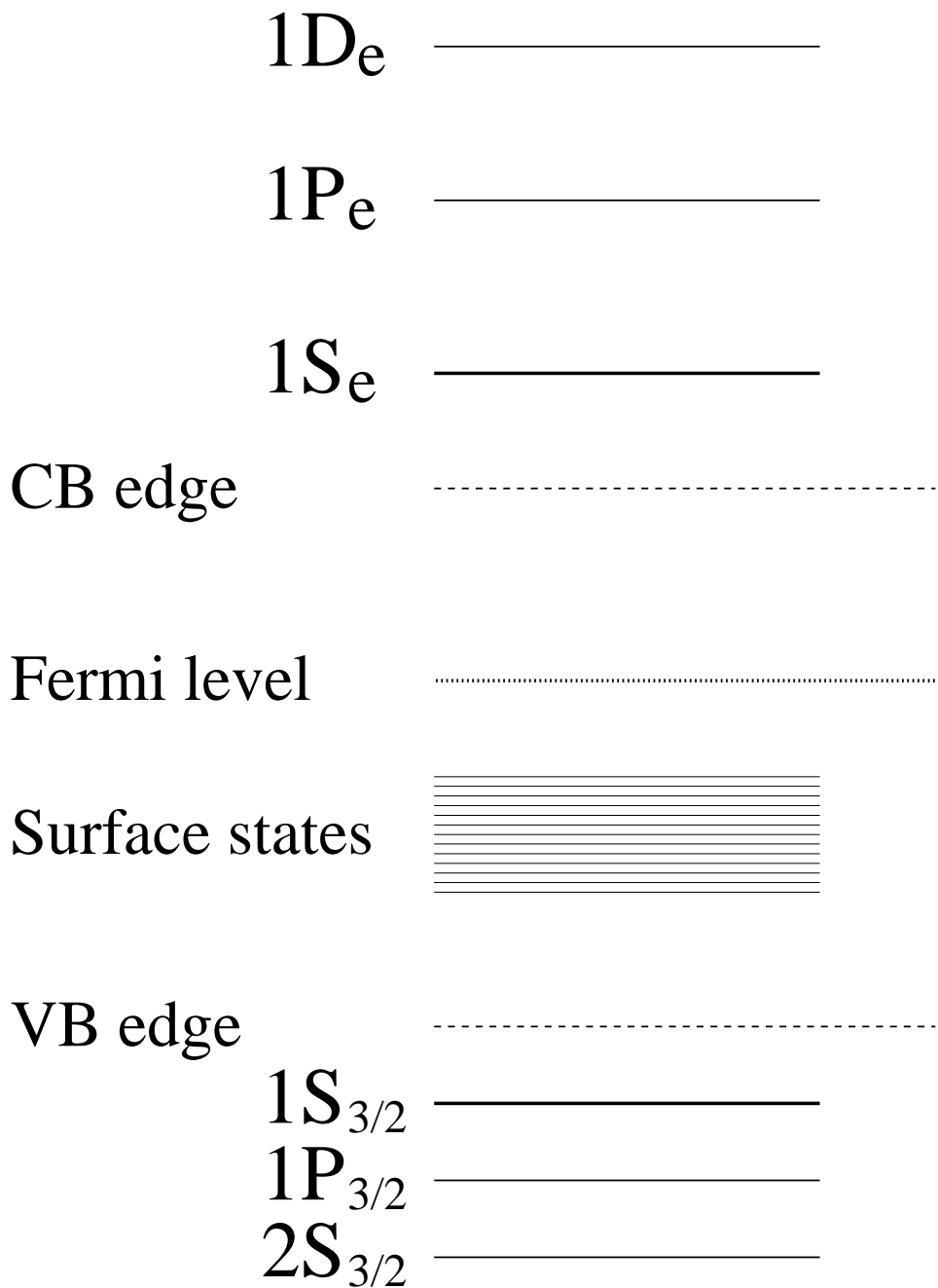

FIG. 1. A schematic picture of the electronic levels of the CdSe quantum dot. $1S_e$, $1P_e$ and $1D_e$ are the delocalized levels in the conduction band (electron states). $1S_{3/2}$, $1P_{3/2}$ and $2S_{3/2}$ are the delocalized levels in the valence band (hole states). The dashed lines represent the energies of the valence band edge and the conduction band edge in the bulk semiconductor. The dotted line represents the Fermi level.

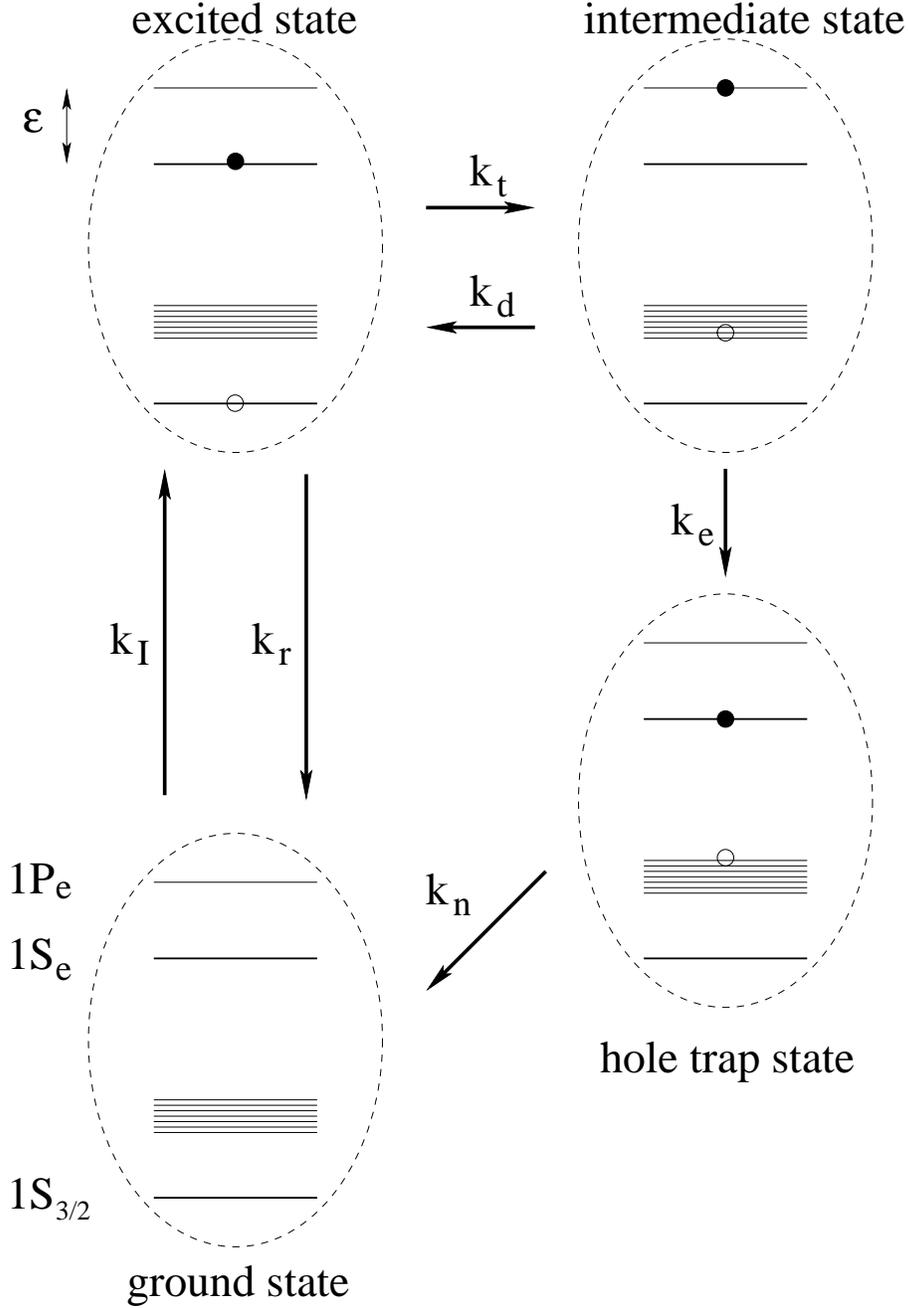

FIG. 2. The kinetic scheme of the QD relaxation after the light-induced excitation

The trapping of the hole is assisted by simultaneous electron 1Se → 1Pe excitation

(Auger process) with consequent relaxation of the excited electron and hole. The energy difference between $1P_e$ state and $1S_e$ state is denoted by $\varepsilon$.

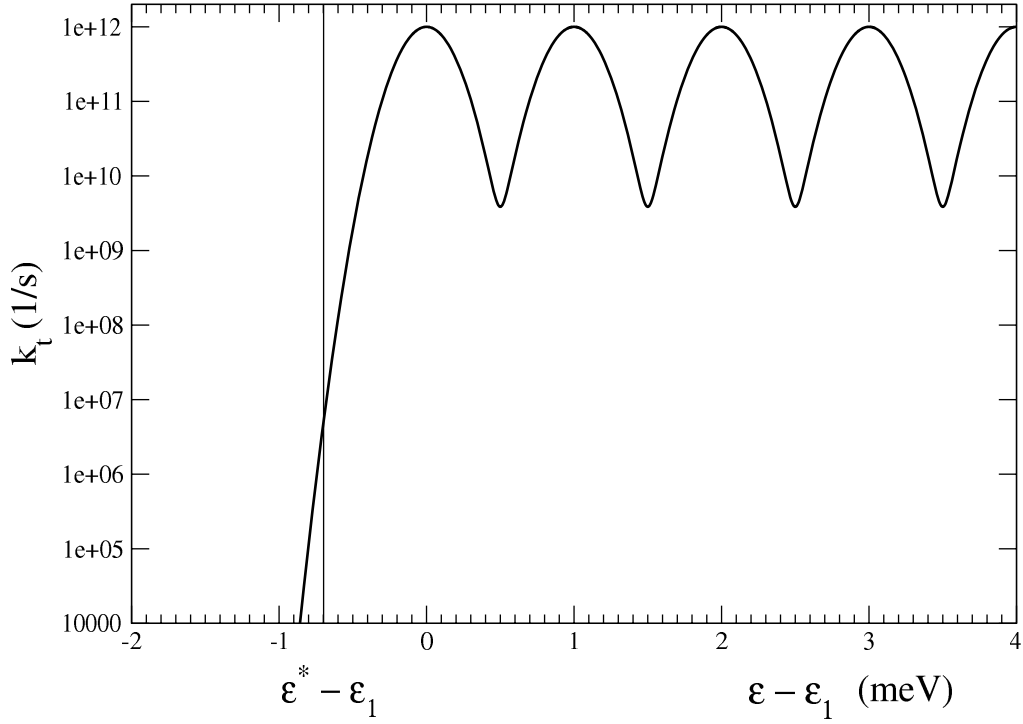

FIG. 3. The trapping reaction rate dependence on the coordinate $\varepsilon$ in the vicinity of $\varepsilon_1$ given by Eq. (2.8). The parameters are: $A_i = 10^{12}$ s$^{-1}$, $\Gamma = 0.2$ meV. The surface states are placed equidistantly $\varepsilon_i = \varepsilon_1 + (i-1) \times 1$ meV.

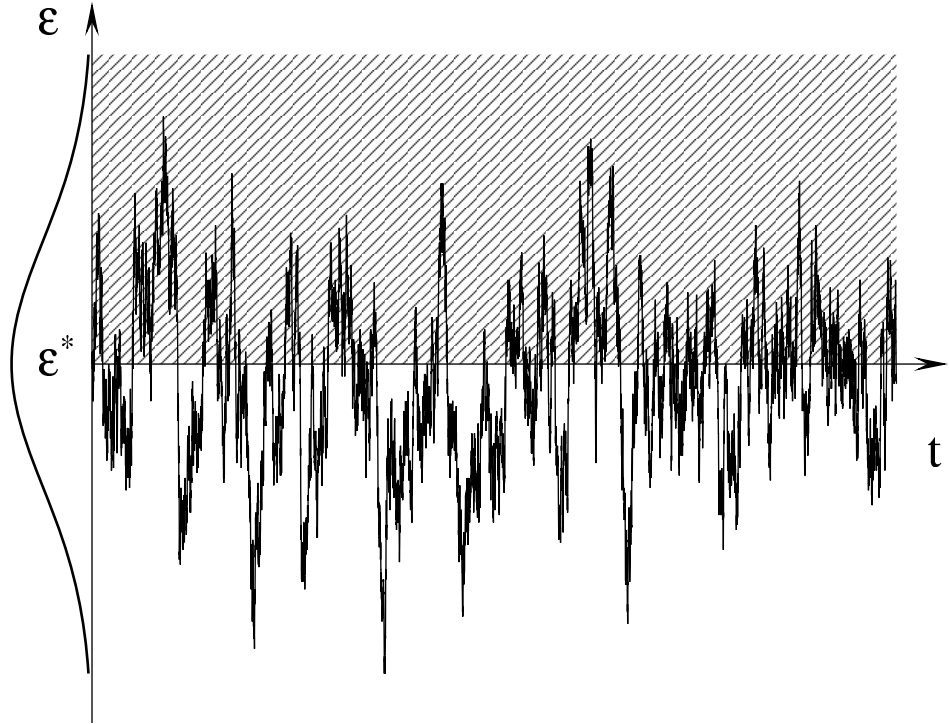

FIG. 4. A schematic picture of the mechanism of the QD blinking. The 1S$_e$-1P$_e$ energy gap $\varepsilon(t)$ is the stochastic Markovian process. The thin curve shows the stationary distribution of $\varepsilon$. The dependence of the QD luminescence intensity on $\varepsilon$ has a sharp threshold at $\varepsilon^*$. So, the luminescence is "off" when $\varepsilon > \varepsilon^*$ (hatched area) and "on" otherwise.

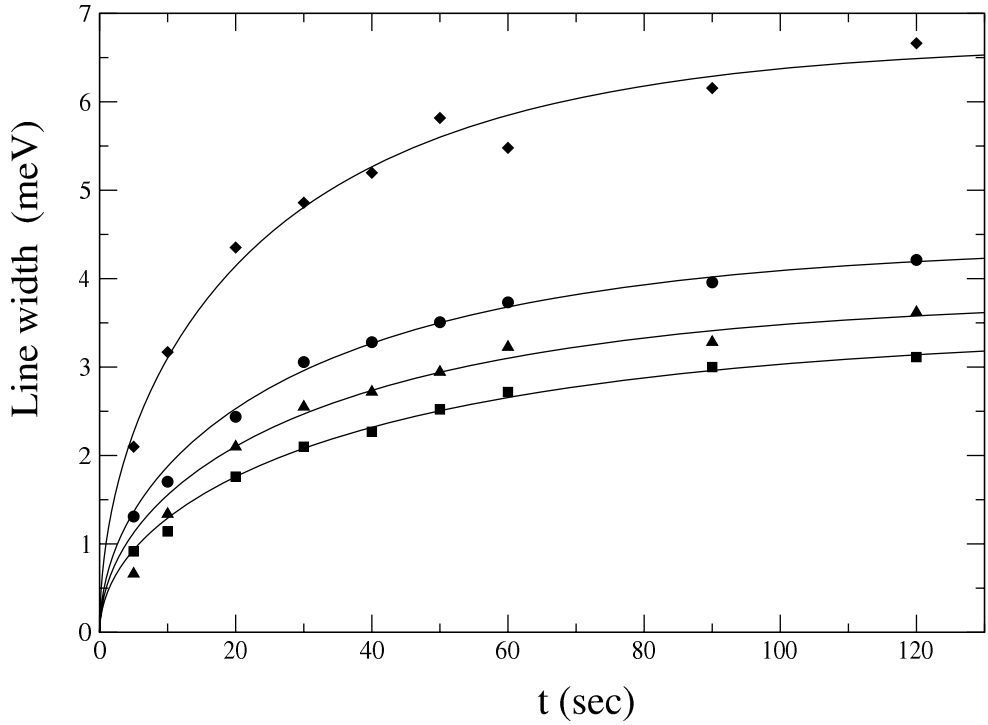

FIG. 5. The averaged single dot spectral line width as a function of the integration time[29] fitted by the formula (4.7). The experimental measurements were taken with the excitation intensity $I = 85$ W/cm$^2$. The parameters of the fits are: $\Delta_E$=1.02 meV, $\tau_E$=130 s, for 10 K (squares); $\Delta_E$=1.6 meV, $\tau_E$=109 s, for 20 K (triangles); $\Delta_E$=1.8 meV, $\tau_E$=99 s, for 30 K (circles); $\Delta_E$=2.8 meV, $\tau_E$=82 s, for 40 K (squares).